\documentclass[prl,twocolumn,showpacs]{revtex4}

\usepackage{graphicx}
\usepackage{amsmath}

\begin{document}

\title{Molecules of Fermionic Atoms in an Optical Lattice}

\author{Thilo St{\"o}ferle, Henning Moritz, Kenneth G{\"u}nter, Michael K{\"o}hl$^*$, and
Tilman Esslinger}

\affiliation{Institute of Quantum Electronics, ETH Z\"{u}rich,
H\"{o}nggerberg, CH--8093 Z\"{u}rich, Switzerland}

\date{\today}

\begin{abstract}

We create molecules from fermionic atoms in a three-dimensional
optical lattice using a Feshbach resonance. In the limit of low
tunnelling, the individual wells can be regarded as independent
three-dimensional harmonic oscillators. The measured binding
energies for varying scattering length agree excellently with the
theoretical prediction for two interacting atoms in a harmonic
oscillator. We demonstrate that the formation of molecules can be
used to measure the occupancy of the lattice and perform
thermometry.
\end{abstract}


\maketitle

Quantum degenerate atomic gases trapped in the periodic potential
of an optical lattice form a quantum many-body system of
unprecedented purity. The short-range interaction due to atom-atom
collisions makes optical lattices ideal to experimentally realize
Hubbard models \cite{Jaksch1998,Hofstetter2002}. Experimental
studies of bosonic Mott insulators
\cite{Greiner2002a,Stoeferle2004,Xu2005,Ryu2005} and of a
fermionic band insulator \cite{Koehl2005} have provided a first
taste of this new approach to quantum many-body physics.

In the vicinity of a Feshbach resonance the collisional
interaction strength between two atoms is tunable over a wide
range. For two fermionic atoms on one lattice site strong
interactions change the properties of the system qualitatively and
physics beyond the standard Hubbard model becomes accessible.
Crossing the Feshbach resonance in one direction leads to an
interaction induced coupling between Bloch bands which has been
observed experimentally \cite{Koehl2005} and described
theoretically \cite{Diener2005}. Crossing the resonance in the
other direction converts fermionic atoms into bosonic molecules.
These processes have no counterpart in standard condensed matter
systems and demand novel approaches to understand the mixed world
of fermions and bosons in optical lattices
\cite{Dickerscheid2005,Carr2005,Zhou2005,Duan2005}. Descriptions
based on multi-band Hubbard models are extremely difficult to
handle and therefore the low-tunneling limit is often used as an
approximation. In this limit the lattice is considered as an array
of microscopic harmonic traps each occupied with two interacting
atoms in different spin states.

The harmonic oscillator with two interacting atoms has been
studied theoretically and the eigenenergies have been calculated
in various approximations
\cite{Busch1998,Blume2002,Bolda2002,Dickerscheid2005}. Its physics
is governed by several length scales. The shortest scale is the
characteristic length of the van-der-Waals interaction potential
between the atoms. The next larger length scale is given by the
s-wave scattering length characterizing low-energy atomic
collisions. However, near the Feshbach resonance it may become
much larger than the extension of the harmonic oscillator ground
state. A precise understanding of the interactions in this
elementary model is a prerequisite in order to comprehend the
many-body physics occurring in optical lattice systems with
resonantly enhanced interactions.

In this paper we study a spin-mixture of fermionic atoms in an
optical lattice and their conversion into molecules by means of a
Feshbach resonance. The binding energy as a function of the s-wave
scattering length between the particles is measured and compared
with theoretical predictions. Moreover, we demonstrate that the
molecule formation can serve as a measure of the temperature of
the atoms in the lattice. For mapping out the phase diagram of
many-body quantum states in the lattice the temperature is a key
parameter. So far, temperature has not been measured in a lattice
since standard methods -- such as observing the rounding-off of
the Fermi surface -- turned out to be dominated by the
inhomogeneity of the trapping potential rather than by temperature
\cite{Rigol2004}. However, we find that the occupancy of the
lattice depends strongly on the temperature
\cite{Katzgraber2005,Koehl2005b,Pupillo2004} and the conversion of
pairs of atoms into molecules is a sensitive probe, similar to the
case of harmonically trapped fermions \cite{Hodby2005}.


In a previous experiment, deeply bound molecules of bosonic atoms
have been created in an optical lattice by photo-association and
were detected by a loss of atoms \cite{Rom2004,Ryu2005}. There the
binding energy is only determined by the atomic properties and
does not depend on the external potential, nor can the scattering
properties between the molecules be adjusted. In low-dimensional
systems molecules produced by a Feshbach resonance using fermionic
atoms have been observed recently \cite{Jochim2003,Moritz2005}.

Our experimental procedure used to produce a degenerate Fermi gas
has been described in detail in previous work \cite{Koehl2005}. In
brief, fermionic $^{40}$K atoms are sympathetically cooled by
thermal contact with bosonic $^{87}$Rb atoms, the latter being
subjected to forced microwave evaporation. After reaching quantum
degeneracy for both species we remove all rubidium atoms from the
trap. The potassium atoms are then transferred from the magnetic
trap into a crossed-beam optical dipole trap which consists of two
horizontally intersecting laser beams. In the optical trap we
prepare a mixture of the $|F=9/2, m_F=-9/2\rangle$ and the
$|F=9/2, m_F=-7/2\rangle$ state with $(50 \pm 4)\%$ in each spin
state \cite{mf} and perform additional evaporative cooling at a
magnetic field of $B=227$\,G. At the end of the evaporation we
reach temperatures of $T/T_F=0.25$ with up to $2 \times 10^5$
particles. The temperatures are determined from a fit to the
density distribution of the noninteracting atomic cloud after
ballistic expansion.

We tune the magnetic field to $B=(210.0 \pm 0.1)$\,G, such that
the s-wave scattering length between the two states vanishes. The
atoms are then transferred into the optical lattice formed by
three orthogonal standing waves with a laser wavelength of
$\lambda=826$\,nm \cite{Koehl2005}. The resulting optical
potential depth $V_0$ is proportional to the laser intensity and
is conveniently expressed in terms of the recoil energy
$E_r=\hbar^2 k^2/(2m)$, with $k=2 \pi / \lambda$ and $m$ being the
atomic mass.

\begin{figure}[htbp]
  \includegraphics[width=.8\columnwidth,clip=true]{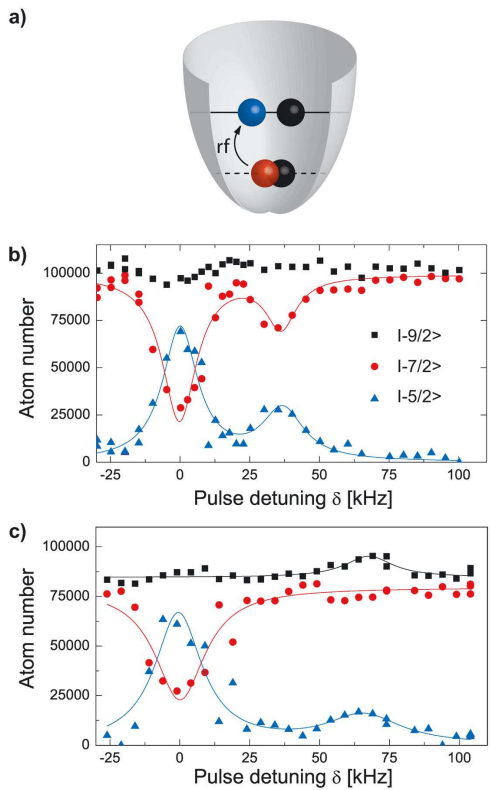}
  \caption{{\bf a)} Illustration of the rf spectroscopy between two bound states
  within a single well of the optical lattice. The atoms in the initial states $|-7/2\rangle$ and
  $|-9/2\rangle$ are converted into a bound dimer by sweeping
  across a Feshbach resonance. Subsequently we drive an rf
  transition $|-7/2\rangle \rightarrow |-5/2\rangle$ to
  dissociate the molecule.
  {\bf b)} rf spectrum taken at $B=202.9$\,G, i.e. for $a<0$.
  {\bf c)} rf spectrum taken at $B=202.0$\,G, i.e. for $a>0$. All data are taken for a lattice depth of 22\,$E_r$.
  The lines are Lorentzian fits to the data.}
  \label{fig1}
\end{figure}

In the optical lattice we create a band insulator for each of the
two fermionic spin states \cite{Koehl2005}. Subsequently, the
molecules are formed by ramping the magnetic field from the zero
crossing of the scattering length at $B=210$\,G in 10 ms to its
final value close to the Feshbach resonance located at
$B_0=202.1$\,G \cite{Regal2004}. From the parameters of our
magnetic field ramp we estimate that the molecule formation is
performed adiabatically. We measure the binding energy $E_B$ of
the dimers by radio-frequency spectroscopy
\cite{Regal2003molecules,Chin2004,Moritz2005}. A pulse with a
frequency $\nu_{RF}$ and a duration of $40\,\mu s$ dissociates the
molecules and transfers atoms from the state $|-7/2\rangle$ into
the initially unpopulated state $|-5/2\rangle$ which does not
exhibit a Feshbach resonance with the state $|-9/2\rangle$ at this
magnetic field. Therefore the fragments after dissociation are
essentially noninteracting (see fig. \ref{fig1}a). We vary the
detuning $\delta=\nu_{RF}-\nu_{0}$ from the resonance frequency
$\nu_0$ of the atomic $|-7/2\rangle \rightarrow |-5/2\rangle$
transition. The power and duration of the pulse is chosen to
constitute approximately a $\pi$-pulse on the free atom
transition. The number of atoms in each spin state is determined
using absorption imaging after ballistic expansion. For this we
ramp down the lattice exponentially with a duration of 1\,ms and
time constant of 0.5\,ms from the initial depth $V_0$ to $5\,E_r$
to reduce the kinetic energy of the gas and then quickly turn off
the trapping potential within a few $\mu$s. The magnetic offset
field is switched off at the start of the expansion, so that no
molecules can be formed in the short time that it passes the
Feshbach resonance. We apply a magnetic field gradient during
3\,ms of the total 7 ms of ballistic expansion to spatially
separate the spin components.

Figure \ref{fig1}b and c show rf spectra of atoms and molecules
trapped in a three-dimensional lattice with a potential depth of
of $V_0=22\,E_r$, which corresponds to a trapping frequency of
$\omega=2 \pi \times 65$\,kHz in the potential wells of the
lattice. The spectrum in figure \ref{fig1}b is taken at a magnetic
field of $B=202.9$\,G, corresponding to $a/a_{ho}=-1.3$. We have
calculated the ground state radius $a_{ho}=64$\,nm by minimizing
the energy of a Gaussian trial wave function inside a single well
of our lattice potential. As compared to a Taylor expansion of the
sinusoidal potential around the minimum this results in slightly
more accurate values. The s-wave scattering length is denoted by
$a$. For negative scattering length the molecules are only bound
when they are strongly confined whereas no bound state would exist
in the homogeneous case. The spectrum exhibits two resonances: the
one at $\delta=0$ corresponds to the atomic transition from the
$|-7/2\rangle$ into the $|-5/2\rangle$ state. This transition
takes place at all lattice sites which initially were only singly
occupied and no molecule has been formed. The second resonance at
$\delta>0$ corresponds to the molecular dissociation and is
shifted from the atomic resonance by the binding energy.
Simultaneous with the increase in the $|-5/2\rangle$ atom number
we observe a loss of atoms in the $|-7/2\rangle$ state, whereas
the $|-9/2\rangle$ remains unaffected.  This is expected since the
rampdown of the lattice before detection dissociates all molecules
and the $|-9/2\rangle$ atom number should be fully recovered.
Residual fluctuations of the $|-9/2\rangle$ atom number are
probably due to uncertainties in the atom number determination
since we do not observe a specific pattern in our spectra.

With the magnetic field tuned to $B=202.0$\,G (see fig.
\ref{fig1}c), which corresponds to $a/a_{ho}=16.8$, the spectrum
changes qualitatively. For this value of the scattering length
stable molecules exist even in free space but the molecules formed
in the lattice are not detected by our state-selective imaging
procedure unless they are dissociated by the rf-pulse. Therefore
only rf dissociated atom pairs show up in the time-of-flight
images, resulting in an increasing number of atoms in the
$|-5/2\rangle$ and the $|-9/2\rangle$ state at the molecular
resonance.

In contrast to earlier work, where molecules were dissociated into
a continuum and the fragments were essentially free particles
\cite{Regal2003molecules,Chin2004,Moritz2005}, the fragments in
our configuration occupy an energy eigenstate of the confining
potential. In such a bound-bound transition no extra kinetic
energy is imparted onto the dissociated fragments since any excess
excitation energy would have to match the band gap. We determine
the binding energy from the separation of the atomic and the
molecular peak. Moreover, since there is at most one molecule
present per lattice site, collisional shifts
\cite{Harber2002,Gupta2003} are absent and we can estimate the
error in the binding energy from the fit error which is less than
5\,kHz.


\begin{figure}[htbp]
  \includegraphics[width=.85\columnwidth,clip=true]{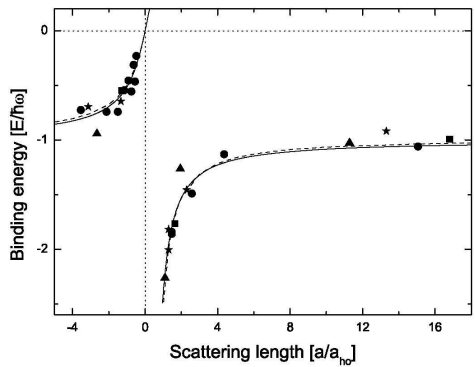}
  \caption{The measured binding energy of molecules in a three-dimensional optical lattice. The data are taken for
  several potential depths of the optical lattice of 6\,$E_r$ (triangles), 10\,$E_r$ (stars), 15\,$E_r$ (circles), and 22\,$E_r$
  (squares). The solid line corresponds to the theory of ref. \cite{Busch1998} with no free
  parameters, the dashed line uses an energy-dependent
  pseudopotential according to ref. \cite{Blume2002}.
  At the position of the Feshbach resonance ($a\rightarrow \pm \infty$)
  the binding energy takes the value $E=-\hbar \omega$.}
  \label{fig2}
\end{figure}

We have investigated the dependence of the binding energy of the
molecules on the scattering length (Fig.\;\ref{fig2}). The
scattering length is derived from the magnetic field using the
parametrization of the Feshbach resonance
$a(B)=a_{bg}(1-\frac{\Delta B}{B-B_0})$, with $a_{bg}=174\,a_0$
\cite{Regal2003background} and $\Delta B=7.8$\,G
\cite{Regal2004lifetime}. We compare our data with the theory for
two particles trapped in a harmonic oscillator potential
interacting via an energy-independent pseudopotential
\cite{Busch1998}. The binding energy $E$ of the molecules depends
on the scattering length according to
\begin{equation}
\sqrt{2} \frac{\Gamma(-E/2\hbar \omega)}{\Gamma(-E/2 \hbar
\omega-1/2)}=\frac{a_{ho}}{a}, \label{Ebinding}
\end{equation}
where $\Gamma(x)$ denotes the Gamma function.  We find the
normalized binding energy $E/\hbar \omega$ to be independent of
the strength of the lattice and all data points agree well with
the theoretical prediction of equation (1) without adjustable
parameters. A pseudopotential approximation is valid, as long as
$a_{ho}$ is large compared to the characteristic length scale of
the van-der-Waals potential between the two atoms $\beta_6=(m
C_6/\hbar^2)^{1/4}$ \cite{Blume2002,Bolda2002}, which for our
experiments is $a_{ho}/\beta_6>10$. However, for $a\gg a_{ho}$ an
energy-dependent pseudopotential will model the system more
accurately. We have calculated the effective range of the
interaction to be $r_{\textrm{eff}}=98$\,a$_0$ \cite{Gao1998} and
the eigenenergies using an energy-dependent pseudopotential
\cite{Blume2002} (dashed line in figure \ref{fig2}). Both models
agree to within a few percent, which is small compared to
experimental uncertainties. Further improvements taking into
account more details of the atom-atom interaction in a two-channel
model have been suggested \cite{Bolda2002,Dickerscheid2005} and
could be tested with our data.

\begin{figure}[htbp]
  \includegraphics[width=.85\columnwidth,clip=true]{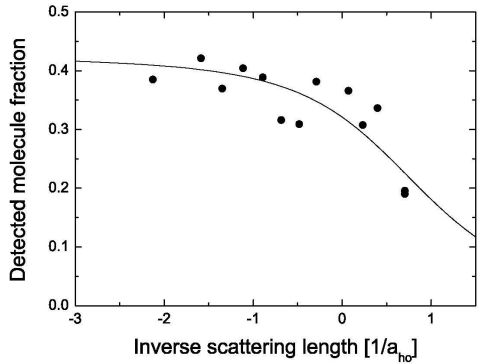}
  \caption{The fraction of molecules detected by rf-spectroscopy at a
  potential depth of $15\,E_r$. For weakly bound molecules ($a_{ho}/a <-1$)
  the dissociation works well since the overlap between the molecular and the atomic wave function is large. For
  deeply bound molecules $a_{ho}/a > 0$ the detected molecular fraction is suppressed due to the vanishing
  overlap between the wave functions. The solid line shows the theoretical
  expectation for a constant molecular fraction of $43\%$.}
  \label{fig3}
\end{figure}

From a quantitative analysis of the spectra we obtain information
about the occupancy of our lattice. We measure the ratio between
the atomic and the molecular peak heights in the spectra of the
$|-5/2\rangle$ atoms and determine the fraction of atoms that
where bound in a molecule. Figure \ref{fig3} shows the measured
data for a lattice with a potential depth of $15\,E_r$. The
detected molecular fraction decreases for large values of $1/a$,
i.e. for deeply bound molecules, because of the small overlap of
the initial molecular and the final atomic wave function in the rf
transition. From the analytical wave function of the molecular
state in relative coordinates $\psi_m({\bf r})$ \cite{Busch1998}
we calculate the overlap integral with the harmonic oscillator
ground state $\psi_{ho}({\bf r})=(\pi a_{ho}^2)^{-3/4} \exp(-{\bf
r}^2/2 a_{ho}^2)$ which determines the relative strength of the
molecular transition assuming that the center-of-mass motion
remains unaffected by the rf transition. We fit the overlap
integral to our experimental data and obtain the fraction of
molecules in the lattice to be $(43\pm 5)\%$ (solid line in figure
\ref{fig3}).

The observed fraction of molecules is primarily determined by the
filling of the lattice. To study the relation between the
temperature and filling we have numerically calculated the density
of states for noninteracting fermions in an optical lattice
including the Gaussian confining potential due to the transverse
envelope of the lattice lasers. In the low-tunnelling limit we
find that the density of states approaches $\rho(E) \propto
E^{\nu}$ with $\nu=1/2$ independent of the lattice depth
\cite{Koehl2005b}. The fraction of doubly occupied lattice sites
in a 50:50 spin mixture depends on $\nu$ and $T/T_F$. This makes
the molecule fraction a quantity ideally suited for thermometry.
Assuming that the occupation probability per spin state and
lattice site $i$ is $0<n_i<1$, the probability of finding two
atoms with different spin state on a lattice site is $n_i^2$. We
identify the molecule fraction with the mean value of $n_i^2$ over
the whole lattice. From a comparison with a numerical calculation
\cite{Koehl2005b} we can conclude that the temperature of the
atoms in the optical lattice is at most $T/T_F=0.45\pm0.03$. A
similar result for our experimental parameters was computed in
\cite{Katzgraber2005}. This value gives an upper limit to the
temperature since it assumes adiabatic formation of molecules at
all doubly occupied sites and a perfect 50:50 mixture of the spin
states. During the ramp across the Feshbach resonance the density
distribution might slightly change as compared to the initial
noninteracting case, which limits the accuracy of the temperature
determination.

We measure the lifetime of the molecules in the lattice to be on
the same order of magnitude as of molecules in an optical dipole
trap \cite{Regal2004lifetime}. This is probably related to the
tunnelling rate, which for our lattice parameters is on the order
of 100\,Hz and therefore the rate of three-body collisions is
comparable to the optical trap. We expect that the lifetime of the
molecules will significantly increase if the lattice depth becomes
larger than 30\,$E_r$. Another possible mechanism which could lead
to a loss of molecules is photo-association induced by the lattice
laser beams.

In conclusion, we have studied molecules in a three-dimensional
optical lattice. We have measured the binding energies and find
good agreement with the fundamental theoretical model of two
interacting particles in a harmonic potential well. Moreover, we
have measured the filling of the lattice by determining the
fraction of molecules formed. This allows for thermometry in the
lattice, which has previously been unaccessible. The fraction of
created molecules gives direct access to the number of doubly
occupied lattice sites in a two-component Fermi gas. Therefore it
could be employed to characterize a Mott insulating phase where
the double occupancy should be strongly reduced.

We would like to thank R. Diener, H. G. Katzgraber, and M. Troyer
for insightful discussions, and OLAQUI, SNF and SEP Information
Sciences for funding.


\begin{thebibliography}{}
\bibitem{mf} In the following we will abbreviate the internal
states of the atom only be the $m_F$ quantum number.
\end{thebibliography}
\end{document}